\documentclass[reqno,a4paper,final,12pt]{amsart}
\usepackage[utf8x]{inputenc}
\usepackage[a4paper,hmargin=2cm,tmargin=3cm,bmargin=3cm]{geometry}
\usepackage{amssymb,amstext,amscd,amsthm,mathrsfs,eucal}
\usepackage{array}
\usepackage[vcentermath]{youngtab}
\Yboxdim{4pt}
\usepackage{hyperref}
\hypersetup{%
  pdftitle   = {Three lectures on 3-algebras},
  pdfkeywords = {Bagger, Lambert, Gustavsson, Filippov, Nambu,
    Faulkner, 3-Lie, 3-Leibniz, 3-algebras, M2-branes, Chern, Simons,
    supersymmetry, superconformal},
  pdfauthor  = {José Miguel Figueroa-O'Farrill},
  pdfcreator = {\LaTeX\ with package \flqq hyperref\frqq}
}
\PrerenderUnicode{éÉ}
\newcommand{\half}{\tfrac12}

\newcommand{\fg}{\mathfrak{g}}

\newcommand{\fso}{\mathfrak{so}}
\newcommand{\fu}{\mathfrak{u}}

\newcommand{\fsp}{\mathfrak{sp}}
\newcommand{\fsu}{\mathfrak{su}}
\newcommand{\fpsu}{\mathfrak{psu}}
\newcommand{\fgl}{\mathfrak{gl}}
\newcommand{\fsl}{\mathfrak{sl}}
\newcommand{\fosp}{\mathfrak{osp}}

\newcommand{\SO}{\mathrm{SO}}

\renewcommand{\Im}{\mathrm{Im}}
\renewcommand{\Re}{\mathrm{Re}}
\ifx\Sp\UnDeFiNeD
\newcommand{\Sp}{\mathrm{Sp}}
\else
\renewcommand{\Sp}{\mathrm{Sp}}
\fi

\newcommand{\RR}{\mathbb{R}}
\newcommand{\CC}{\mathbb{C}}

\newcommand{\eD}{\mathscr{D}}
\newcommand{\eL}{\mathscr{L}}

\newcommand{\XX}{X}
\newcommand{\YY}{Y}
\newcommand{\XXbar}{\overline{X}}

\DeclareMathOperator{\Tr}{Tr}
\DeclareMathOperator{\End}{End}

\DeclareMathOperator{\ad}{ad}
\DeclareMathOperator{\AdS}{AdS}

\newcommand{\rf}[1]{[\![#1]\!]}
\theoremstyle{plain}
\newtheorem{lemma}{Lemma}

\newtheorem{theorem}[lemma]{Theorem}

\theoremstyle{definition}

\newcommand{\MUNCH}[1]{\relax}

\allowdisplaybreaks[1]
\begin{document}
\title{Three lectures on 3-algebras}
\author{José Miguel Figueroa-O'Farrill}
\address{Maxwell Institute and School of Mathematics, University of Edinburgh}
\address{Departament de Física Teòrica, Universitat de València}
\email{j.m.figueroa@ed.ac.uk}
\begin{abstract}
  These notes are based on lectures given in Valencia in October 2008
  and in Stockholm in November 2008, in the framework of the Nordita
  workshop ``Geometrical aspects of String Theory''.  We introduce the
  notion of a metric 3-Lie algebra and review some of the classification
  results.  We explain the deconstruction of metric 3-Lie algebras in
  Lie algebraic terms and introduce a general framework in which to
  describe other 3-algebras of relevance in the description of
  three-dimensional superconformal Chern--Simons theories, particularly
  those with $N{=}6$.  The emphasis throughout is on the general ideas
  and concrete examples.
\end{abstract}
\maketitle
\tableofcontents

\section{``3 is the new 2''}
\label{sec:metric-3-lie}

In this lecture I will introduce metric 3-Lie algebras\footnote{Until
  recently I used to call them \emph{Lie 3-algebras}, but my
  $n$-categorical friends insist that I should call them \emph{3-Lie
    algebras} instead.  I will nevertheless continue to use
  \emph{3-algebra} for the generic case.}, give some examples and state
a number of classification results.

\subsection{Metric 3-Lie algebras}
\label{sec:metric-3-lie-1}

The Bagger--Lambert--Gustavsson (BLG) proposal
\cite{BL1,GustavssonAlgM2,BL2} for a superconformal field theory dual to
a stack of M2-branes is essentially a maximally supersymmetric
Chern--Simons + matter theory.  In three dimensions this means an
$N{=}8$ theory (i.e., a theory with 16 real supercharges) realising the
Killing superalgebra $\fosp(8|4)$ of the near-horizon geometry $S^7
\times \AdS_4$ of the M2 branes.  This superalgebra has bosonic
subalgebra $\fso(8) \oplus \fso(3,2) \cong \fso(8) \oplus \fsp(4,\RR)$
and odd subspace in their fundamentals, hence the name.  The novel
feature of the BLG model is that the matter fields take values in a
vector space $V$ with a trilinear bracket
\begin{equation}
  \label{eq:3bracket}
    V \times V \times V \to V  \qquad\text{sending}\qquad (x,y,z)
    \mapsto [x,y,z]~,
\end{equation}
and a symmetric inner product
\begin{equation}
  \label{eq:IP}
  V \times V \to \RR \qquad\text{denoted}\qquad  (x,y) \mapsto
  \left<x,y\right>~,
\end{equation}
satisfying a number of identities:
\begin{enumerate}
\item total \emph{skewsymmetry} of the bracket, whence it defines a linear map
  $\Lambda^3 V \to V$;
\item \emph{metricity}:
  \begin{equation}
    \label{eq:metricity}
    \left<[x,y,z_1],z_2\right> = - \left<[x,y,z_2],z_1\right>~;
  \end{equation}
\item and the so-called \emph{fundamental identity}:
  \begin{equation}
    \label{eq:FI}
    [x,y,[z_1,z_2,z_3]] =  [ [x,y,z_1],z_2,z_3] + [z_1, [x,y,z_2],z_3] +  [z_1,z_2,[x,y,z_3]]~,
  \end{equation}
  for all $x,y,z_i \in V$.  (This identity will be rewritten below in
  a more succinct and conceptual manner.)
\end{enumerate}
We will call such a vector space $V$ with the bracket and the inner
product a \textbf{metric 3-Lie algebra}, a concept which --- perhaps
without the metricity assumption --- is due to Filippov \cite{Filippov}.

The first remark is that this is a natural generalisation of the
concept of a metric Lie algebra; which is a vector space $\fg$
together with a bilinear bracket $\fg \times \fg \to \fg$, sending
$(X,Y)$ to $[X,Y]$ and a symmetric inner product $\fg \times \fg \to
\RR$, sending $(X,Y)$ to $\left<X,Y\right>$, obeying the following
identities:
\begin{enumerate}
\item \emph{skewsymmetry} of the bracket, whence it defines a linear map
  $\Lambda^2 \fg \to \fg$;
\item \emph{metricity}:
  \begin{equation}
    \label{eq:metricityLA}
    \left<[X,Y],Z\right> = - \left<[X,Z],Y\right>~;
  \end{equation}
\item and the \emph{Jacobi identity}:
  \begin{equation}
    \label{eq:JI}
    [X,[Y,Z]] =  [[X,Y],Z] + [Y, [X,Z]]~,
  \end{equation}
  for all $X,Y,Z \in \fg$.
\end{enumerate}
Just like the Jacobi identity can be reinterpreted as saying that for
all $X \in \fg$, $\ad_X := [X,-]$ is a derivation over the bracket, the
fundamental identity in a 3-Lie algebra $V$ says that for all $x,y\in
V$, $D(x,y):= [x,y,-]$ is a derivation over the 3-bracket.  In fact, as
we will see, many of the known results for metric Lie algebras hold word
for word (but after some reinterpretation) also for metric 3-Lie
algebras, and in fact even for metric $n$-Lie algebras, for $n>3$,
defined in the obvious way.

The second remark, which is not to be taken too seriously, is that
metric (\emph{2}-)Lie algebras appear prominently in
\emph{two}-dimensional superconformal field theory, via their rôle in
the Sugawara construction; metric \emph{3}-Lie algebras appear
prominently in \emph{three}-dimensional superconformal field theory, via
their rôle in the BLG model; and that the self-dual five-form in $\AdS_5
\times S^5$, a crucial background in much of today's interest on
\emph{four}-dimensional superconformal field theory, defines a metric
\emph{4}-Lie algebra. (While curious, I do not expect this to
generalise.)

\subsection{The Nambu bracket}
\label{sec:nambu-bracket}

Being a geometrical meeting, let us start with a geometrical example
of a metric 3-Lie algebra.  Let $(M,\omega)$ be a compact, oriented
$3$-dimensional manifold and $\omega$ a nowhere-vanishing $3$-form
defining the orientation.  We will also assume that $M$ has no
boundary.  Given three smooth functions $f,g,h \in C^\infty(M)$ the
wedge product of their differentials is a 3-form which must be
proportional to $\omega$.  We define the function $\{f,g,h \} \in
C^\infty(M)$ by
\begin{equation}
  \label{eq:Nambu}
  df \wedge dg \wedge dh = \{f,g,h \} \omega~.
\end{equation}
This defines an alternating trilinear map
\begin{equation}
  \label{eq:Nambu3bracket}
  C^\infty(M) \times   C^\infty(M) \times   C^\infty(M) \to
  C^\infty(M)~,
\end{equation}
sending $(f,g,h)$ to $\{f,g,h\}$.  This bracket, originally due to
Nambu \cite{Nambu:1973qe}, obeys the following properties:
\begin{enumerate}
\item Leibniz rule:
  \begin{equation}
    \label{eq:LeibnizRule}
    \{f,g,h_1 h_2\}  = \{f,g,h_1\} h_2 + h_1 \{f,g, h_2\}~,
  \end{equation}
\item the \emph{fundamental identity}:
  \begin{equation}
    \label{eq:NambuFI}
    \{f,g,\{h_1,h_2,h_3\}  \} = \{\{f,g,h_1\},h_2,h_3\} +
    \{h_1,\{f,g,h_2\},h_3\} + \{h_1,h_2,\{f,g,h_3\}\}~,
  \end{equation}
  for all $f,g,h_i \in C^\infty(M)$.
\end{enumerate}

Only the last identity requires proof.  We start by observing that the
Leibniz rule says that given $f,g\in C^\infty(M)$, the map
$C^\infty(M) \to C^\infty(M)$ sending $h$ to $\{f,g,h\}$ is a
derivation, whence it defines a vector field $X_{f,g}$.  This vector
field leaves $\omega$ invariant, as can be seen by contracting it into
both sides of equation \eqref{eq:Nambu}:
\begin{equation}
  df \wedge dg \{f,g,h\} = \{f,g,h\} \imath_{X_{f,g}}\omega~,
\end{equation}
whence
\begin{equation}
  \imath_{X_{f,g}}\omega = df \wedge dg~,
\end{equation}
and hence
\begin{equation}
  \eL_{X_{f,g}} \omega = d\imath_{X_{f,g}} \omega = d(df \wedge dg) = 0~.
\end{equation}
Now we differentiate
\begin{equation}
  dh_1 \wedge dh_2 \wedge dh_3 = \{h_1,h_2,h_3\} \omega
\end{equation}
with respect to $X_{f,g}$, using that the Lie and exterior derivatives
commute, to obtain
\begin{equation}
  d\{f,g,h_1\} \wedge dh_2 \wedge dh_3 +   dh_1 \wedge d\{f,g,h_2\}
  \wedge dh_3 + dh_1 \wedge dh_2 \wedge d\{f,g,h_3\}  =
\{f,g,\{h_1,h_2,h_3\}\} \omega~,
\end{equation}
which, using equation \eqref{eq:Nambu} again on the three terms in the
left-hand side, becomes equation \eqref{eq:NambuFI}.

Furthermore if we define an inner product on $C^\infty(M)$ by
\begin{equation}
  \label{eq:NambuIP}
  \left<f,g\right> := \int_M f g \omega~,
\end{equation}
for all $f,g\in C^\infty(M)$, then we have that
\begin{equation}
  \label{eq:Nambumetricity}
  \left<\{f,g,h_1\},h_2\right> = - \left<\{f,g,h_2\},h_1\right>~,
\end{equation}
for all $f,g,h_i \in C^\infty(M)$.  Indeed,
\begin{align*}
  \left<\{f,g,h_1\},h_2\right> &= \int_M \{f,g,h_1\} h_2 \omega\\
  &= \int_M df \wedge dg \wedge dh_1 h_2\\
  &= \int_M df \wedge dg \wedge d(h_1h_2) - \int_M df \wedge dg \wedge
  dh_2 h_1\\
  &= \int_M d( h_1 h_2 df \wedge dg) - \int_M \{f,g,h_2\} h_1 \omega\\
  &= - \left<\{f,g,h_2\},h_1\right>~,
\end{align*}
since $M$ has no boundary.

\subsection{Structure of metric 3-Lie algebras}
\label{sec:structure-3-lie}

The Jacobi identity \eqref{eq:JI} for a Lie algebra $\fg$ is
equivalent to $[\ad_X,\ad_Y] = \ad_{[X,Y]}$ in $\End\fg$, for all
$X,Y\in\fg$.  Now we will reinterpret the fundamental identity
\eqref{eq:FI} of a 3-Lie algebra $V$ in a similar way.  Let us define
$D: \Lambda^2 V \to \End V$ by extending
\begin{equation}
  D(x\wedge y) = [x,y,-]
\end{equation}
linearly from monomials to arbitrary elements of $\Lambda^2 V$.  Then
the fundamental identity \eqref{eq:FI} is equivalent to
\begin{equation}
  \label{eq:FID}
  [D(X),D(Y)] = D(D(X)\cdot Y) \qquad\text{for all $X,Y\in \Lambda^2V$,}
\end{equation}
where the bracket on the left-hand side is the commutator in $\End V$
and the $\cdot$ in the right-hand side is the natural action of $\End
V$ on $\Lambda^2 V$; i.e.,
\begin{equation}
  \label{eq:EndVaction}
  D(X) \cdot (x \wedge y) = D(X) \cdot x \wedge y + x \wedge D(X)
  \cdot y~.
\end{equation}
Indeed, if we now take $X = x \wedge y$ and $Y = z_1\wedge z_2$ and we
apply \eqref{eq:FID} on $z_3\in V$, we obtain the fundamental identity
\eqref{eq:FI} upon using \eqref{eq:EndVaction}.

The fundamental identity in the form \eqref{eq:FID} says that the
image of $D$ is a Lie subalgebra of $\fgl(V)$, or indeed $\fso(V)$ if
metric.  In the case of a metric Lie algebra, the Jacobi identity says
that $\ad : \fg \to \fso(\fg)$ is a Lie algebra homomorphism.  But
what about in the case of (metric) 3-Lie algebras?  By analogy with
Lie algebras we could define a bracket on $\Lambda^2 V$ by
\begin{equation}
  [X,Y] := D(X) \cdot Y~,
\end{equation}
in terms of which, the fundamental identity \eqref{eq:FID} applied to
$Z \in \Lambda^2 V$ becomes a version of the Jacobi identity for a Lie
algebra:
\begin{equation}
  \label{eq:LI}
  [X,[Y,Z]] = [[X,Y],Z] + [Y,[X,Z]]~.
\end{equation}
Since in general $[X,Y] \neq - [Y,X]$, $\Lambda^2 V$ is not a Lie
algebra but only a \textbf{(left) Leibniz algebra} --- a sort of
noncommutative version of a Lie algebra, introduced by Loday in
\cite{MR1217970} and much studied since.  Nevertheless, the map $D$ is
still a Leibniz algebra homomorphism.  (Notice that a Lie algebra is in
particular also a Leibniz algebra.)  I will not say more about Leibniz
algebras here, except to note that they underlie many of the structural
(an in particular cohomological) properties of 3-Lie algebras (and their
relatives).  For instance, the deformation theory of a 3-Lie algebra $V$
is governed by the cohomology of its associated Leibniz algebra
$\Lambda^2 V$.  Finally, let me point out that the correspondence from
3-Lie algebras to Leibniz algebras sending $V$ to $\Lambda^2 V$ is
functorial.

Given the similarity between 3-Lie algebras and Lie algebras, it is
worth contrasting the two.  There are many question one can ask and most
have already been answered.  We point out three references: the original
paper of Filippov \cite{Filippov}, a later paper of Kasymov
\cite{Kasymov} and the PhD thesis of Ling \cite{LingSimple}.  In these
papers there is already a well-developed structure theory of 3-Lie
algebras with all the usual concepts (often refined): ideals and
homomorphisms, nilpotency, solvability, radical,... There is even a
Levi-Malcev theorem stating that, just as for Lie algebras, a 3-Lie
algebra is a semidirect product of a semisimple 3-Lie algebra and a
solvable 3-Lie algebra (its radical).  Just as for Lie algebras, a
semisimple 3-Lie algebra is a direct sum of simple ideals, but
\emph{unlike} in the case of Lie algebras, where there are infinite
isomorphism classes of simple Lie algebras, there is over the complex
numbers a \emph{unique simple} 3-Lie algebra: $V=\CC^4$ with basis
$(e_1,\dots, e_4)$ and bracket
\begin{equation}
  [e_i,e_j,e_k] = \varepsilon_{ijk\ell} e_\ell~,
\end{equation}
where the Levi-Cività symbol is normalised to $\varepsilon_{1234} =
1$.  This result is proved in \cite{LingSimple}, also for $n$-Lie
algebras with $n>3$.  Over the real numbers, we simply attach signs
$\sigma_i$ to the right-hand side of the bracket:
\begin{equation}
  [e_i,e_j,e_k] = \varepsilon_{ijk\ell} \sigma_\ell e_\ell~.
\end{equation}
It is an easy exercise to show that this defines on $\RR^4$ the
structure of a metric 3-Lie algebra relative to the inner product
\begin{equation}
  \left<e_i,e_j\right> = \sigma_i \delta_{ij}~.
\end{equation}
Taking all $\sigma_i = 1$, we obtain a 4-dimensional euclidean metric
3-Lie algebra, which we denote $A_4$.  Similarly, we can define
$A_{3,1}$ and $A_{2,2}$ by changing the sign of one or two of the
$\sigma_i$ and in this way obtain lorentzian and split metric 3-Lie
algebras, respectively.  It turns out that $A_4$ is the unique
nonabelian indecomposable such metric 3-Lie algebra, a result
conjectured in \cite{FOPPluecker}, and proved independently in
\cite{NagykLie,GP3Lie,GG3Lie}.  As discussed in \cite{Lor3Lie}, it also
follows easily from the classification of simple 3-Lie algebras.  By way
of contrast, the metric Lie algebras admitting a positive-definite
invariant inner product are the reductive Lie algebras, which are direct
sums of semisimple and abelian.  Of course, the same is true for 3-Lie
algebras (and indeed for $n$-Lie algebras, $n>3$), except that there is
a unique such simple object.

Leaving questions of manifest unitarity of the lagrangian aside, every
metric 3-Lie algebra (of any signature) gives rise to a
three-dimensional $N{=}8$ supersymmetric Chern--Simons theory with
matter.  This suggests that the classification of metric 3-Lie algebras
is an interesting problem.  If only to temper our expectations, one can
ask what is known about metric Lie algebras.

Given two metric Lie algebras one can take their orthogonal direct sum
to construct another metric Lie algebra.  We say that a metric Lie
algebra is \textbf{indecomposable} if it cannot be written in this
way.  It is therefore only necessary to classify the indecomposable
ones.  The same holds for 3-Lie algebras, and indeed for $n$-Lie
algebras for $n>3$.

There is no classification of metric Lie algebras beyond index 3.  (I
recall that the index of an inner product of signature $(p,q)$ is the
minimum of $p$ and $q$, whence index $0$ is (by convention)
positive-definite, index $1$ is lorentzian,...).  The case of index 0
is classical: the indecomposable objects are the compact simple Lie
algebras and $\fu(1)$.  The case of index 1 is due to Medina
\cite{MedinaLorentzian} and the cases of index 2 and 3 due to Kath and
Olbrich \cite{KathOlbrich2p,MR2205075}.  For later comparison, here is
the statement for the lorentzian case.

\begin{theorem}
  Every (finite-dimensional) lorentzian Lie algebra $\fg$ is the
  orthogonal direct sum $\fg = \fg_0 \oplus \fg_1$, where $\fg_0$ is
  an indecomposable lorentzian Lie algebra and $\fg_0$ is a
  positive-definite (hence, reductive) Lie algebra.  A
  finite-dimensional indecomposable lorentzian Lie algebra is
  isomorphic to one of the following:
  \begin{enumerate}
  \item one-dimensional with negative-definite inner product;
  \item $\fsl(2,\RR)$ with respect to (the negative of) the Killing
    form; or
  \item $E \oplus \RR u \oplus \RR v$, where $E$ is
    an even-dimensional euclidean vector space, and the inner product 
    $\left<-,-\right>$ extends that of $E$ by declaring $u,v \perp E$,
    $\left<u,v\right>=1$ and $\left<v,v\right>=0$, and the Lie
    brackets are given by
    \begin{equation*}
      [u,x] = J(x) \qquad\text{and}\qquad [x,y] = \left<J(x),y\right>
      v~,
    \end{equation*}
    for all $x,y \in E$ and where $J \in \fso(E)$ is a nondegenerate
    skewsymmetric endomorphism.
  \end{enumerate}
\end{theorem}

The simplest example of the third type is the famous Nappi--Witten Lie
algebra \cite{NW}, where $E=\RR^2$ and $J$ is an orthogonal complex
structure.  It can also be interpreted as a central extension of the
Lie algebra of euclidean motions in two dimensions.

The most general result in the theory of metric Lie algebras is the
structure theorem of Medina and Revoy \cite{MedinaRevoy} (see also
\cite{FSalgebra}), which says that the class of finite-dimensional
metric Lie algebras is generated by the simple and the one-dimensional
Lie algebras using two operations: orthogonal direct sum and
\emph{double extension}.  (We will not define this notion here.)

Based on this brief summary of the state of the art on metric Lie
algebras, we should perhaps not expect to do much better for metric
3-Lie algebras.  In fact, in addition to the index 0 results mentioned
above, there are classifications for lorentzian \cite{Lor3Lie} and index
2 \cite{2p3Lie} 3-Lie algebras.  It should be possible to go further and
classify index-3 3-Lie algebras, but we have not found the need (nor the
energy) to do so.  Moreover there is an identical-sounding structure
theorem for metric 3-Lie algebras \cite{2p3Lie} and for metric $n$-Lie
algebras for $n>3$ \cite{JMFMetricNLie}, except that the notion of
double extension is even more cumbersome to define.

I will finish this first lecture with the statement of the lorentzian
result.  A very similar result holds also for metric $n$-Lie algebras
for $n>3$ \cite{JMFLorNLie}.

\begin{theorem}
  Every (finite-dimensional) lorentzian 3-Lie algebra $V$ is the
  orthogonal direct sum $V = V_0 \oplus V_1$, where $V_0$ is
  an indecomposable lorentzian 3-Lie algebra and $V_1$ is a
  positive-definite 3-Lie algebra.  A finite-dimensional
  indecomposable lorentzian 3-Lie algebra is isomorphic to one of the
  following:
  \begin{enumerate}
  \item one-dimensional with negative-definite inner product;
  \item the simple 3-Lie algebra $A_{3,1}$; or
  \item $\fg \oplus \RR u \oplus \RR v$, where $\fg$ is
    a semisimple Lie algebra with a choice of positive-definite inner
    product $\left<-,-\right>$ which we extend to the whole space by
    declaring $u,v \perp \fg$, $\left<u,v\right>=1$ and
    $\left<v,v\right>=0$, and the 3-brackets are given by
    \begin{equation*}
      [u,x,y] = [x,y] \qquad\text{and}\qquad [x,y,z] =
      -\left<[x,y],z\right> v~,
    \end{equation*}
    for all $x,y,z \in \fg$.
  \end{enumerate}
\end{theorem}

The metric 3-Lie algebras in the third class were discovered
independently in \cite{GMRBL,BRGTV,HIM-M2toD2rev}.

The index-2 classification is detailed in \cite{2p3Lie}, but even just
listing them would already take a lot of space.  In that paper we
repackage some desirable physical properties of the BLG model as
3-algebraic criteria, which can then be applied to further refine the
classification.  This results in two main classes of ``physically
interesting'' index-2 3-Lie algebras, which are the subject of a
forthcoming paper \cite{2pBL}.

\section{``2 strikes back''}
\label{sec:2-strikes-back}

In this second lecture we relate metric 3-Lie algebras and, more
generally, also other metric 3-Leibniz algebras of relevance in
three-dimensional superconformal Chern--Simons theories, to the
representation theory of metric Lie algebras.  This is done by adapting
a general algebraic construction due to Faulkner \cite{FaulknerIdeals},
which we will present at the end of the lecture.  This lecture is based
on \cite{Lie3Algs}.

\subsection{Deconstructing the metric 3-Lie algebras}
\label{sec:deconstruction}

Let $V$ be a metric 3-Lie algebra, defined by a linear map $D:
\Lambda^2 V \to \fso(V)$ obeying the fundamental identity
\eqref{eq:FID}, whence the image $\fg$ of $D$ is a Lie subalgebra of
$\fso(V)$.  The surprising thing is that $\fg$ is a \emph{metric} Lie
algebra relative to the inner product defined by extending
\begin{equation}
  \left(D(x \wedge y), D(u\wedge v)\right) = \left<[x,y,u],v\right>
\end{equation}
linearly to all of $\fg$.  First of all, we notice that the above
expression actually defines a symmetric bilinear form on $\fg$:
indeed, metricity and the skewsymmetry of the 3-bracket, says that
$\left<[x,y,u],v\right>$ is totally skewsymmetric in all four
arguments.  In particular it is symmetric under the interchange of
pairs:
\begin{equation}
   \left(D(x \wedge y), D(u\wedge v)\right)  = \left(D(u\wedge v), D(x
     \wedge y)\right)~.
\end{equation}
Next we show that this bilinear form is non-degenerate.  Let $\delta
\in \fg$ be perpendicular to all of $D(u\wedge v)$.  Then
\begin{align*}
  \left(\delta, D(u,v)\right) = \left<\delta u, v\right> = 0
  \qquad\text{for all $u,v\in V$}~.
\end{align*}
Since the inner product on $V$ is nondegenerate, this means $\delta u
= 0$ for all $u \in V$, whence $\delta$, being an endomorphism, must
vanish.  This says that $\left(-,-\right)$ so defined is a
nondegenerate symmetric bilinear form; that is, an inner product.
Finally, we show that it is invariant.  Let $X,Y,Z = u \wedge v \in
\Lambda^2V$ and consider
\begin{align*}
  \left(D(X),[D(Y),D(Z)]\right) &= \left(D(X),D(D(Y)\cdot Z)\right) &&
  \text{by \eqref{eq:FID}}\\
  &= \left(D(X),D(Y)\cdot u \wedge v + u \wedge D(Y) \cdot v\right)\\
  &= \left<D(X)\cdot D(Y)\cdot u, v\right> + \left<D(X)\cdot u, D(Y)
    \cdot v\right>\\
  &= \left<D(X)\cdot D(Y)\cdot u, v\right> - \left<D(Y) \cdot
    D(X)\cdot u, v\right>\\
  &= \left<[D(X),D(Y)]\cdot u, v\right>\\
  &= \left([D(X),D(Y)],D(Z)\right)~.
\end{align*}
Further we notice that every $D(x\wedge y)$ is null because of the
skewsymmetry of the bracket:
\begin{equation}
  \left(D(x\wedge y), D(x\wedge y\right) = \left<[x,y,x],y\right> =
  0~.
\end{equation}
In particular this means that $\left(-,-\right)$ must have split
signature and hence that $\fg$ is even-dimensional.

For example, for the positive-definite simple 3-Lie algebra $A_4$, all
$D(e_i \wedge e_j)$ are linearly independent for $i<j$,
whence they span all of $\fso(4)$.  The inner product is
\begin{equation}
  \left(D(e_i \wedge e_j),D(e_k \wedge e_\ell)\right) =
  \varepsilon_{ijk\ell}~,
\end{equation}
which is split.  (In fact, it is just given by the wedge product,
under the isomorphism of $\fso$ with $\Lambda^2$.)

In summary, we have managed to deconstruct a metric 3-Lie algebra $V$
into a metric Lie subalgebra $\fg < \fso(V)$.  It is important to
remark that in general, the invariant inner product on $\fg$ need not
be the restriction of an invariant inner product on $\fso(V)$.  A
natural question is then whether one can reconstruct the metric 3-Lie
algebra $V$ from such data; namely a metric Lie algebra $\fg$ and a
faithful orthogonal representation $V$.

\subsection{Reconstructing a metric 3-Leibniz algebra}
\label{sec:reconstr-metr-3}

Let $\fg$, with inner product $\left(-,-\right)$, be a metric Lie
algebra and let $V$ be a faithful orthogonal representation, so that
we can think of $\fg$ as a Lie subalgebra of $\fso(V)$.  Given $x,y
\in V$, we define $D(x,y) \in \fg$ by transposing the $\fg$-action:
\begin{equation}
  \left(D(x,y),X\right) = \left<X\cdot x, y\right> \qquad\text{for all
  $X\in \fg$,}
\end{equation}
where $\cdot$ denotes the $\fg$ action.  This defines $D(x,y)$
uniquely, because $\left(-,-\right)$ is nondegenerate.  We define a
3-bracket on $V$ by
\begin{equation}
  [x,y,z] := D(x,y)\cdot z~,
\end{equation}
for all $x,y,z \in V$.  Since $D(x,y) \in \fso(V)$, we have the
following identity
\begin{align*}
  \left<[x,y,z],w\right> &= \left<D(x,y)\cdot z, w\right> \\
  &= - \left<z,D(x,y)\cdot w\right>\\
  &= - \left<z,[x,y,w]\right>~.
\end{align*}
Also we have that the symmetry of the inner product on $\fg$,
\begin{equation}
  \left(D(x,y),D(u,v)\right)=   \left(D(u,v),D(x,y)\right)
\end{equation}
translates into
\begin{equation}
  \label{eq:symmetry}
  \left<[x,y,u],v\right> = \left<[u,v,x],y\right>~,
\end{equation}
which together with metricity implies that
\begin{equation}
  \label{eq:2skew}
  [x,y,z] = - [y,x,z] \quad \iff \quad D(x,y) = - D(y,x)~,
\end{equation}
whence we may think of $D$ as a linear map $D: \Lambda^2 V \to \fg$.
In addition, the 3-bracket obeys the fundamental identity
\eqref{eq:FID}.  Indeed, for all $X,Y \in \fg$ and $u,v\in V$,
\begin{align*}
  \left([D(u\wedge v),X],Y\right) &= \left(D(u\wedge v),[X,Y]\right)
  && \text{by invariance}\\
  &= \left<[X,Y]\cdot u, v\right>\\
  &= \left<X\cdot Y \cdot u, v\right> - \left<Y\cdot X\cdot u,
    v\right>\\
  &= - \left<Y \cdot u, X\cdot v\right> - \left<Y\cdot X\cdot u,
    v\right>\\
  &= - \left(D(u \wedge X \cdot v), Y\right) - \left(D(X \cdot u
    \wedge v), Y\right)~,
\intertext{whence}
[X,D(u\wedge v)] &= D(X\cdot (u \wedge v))~,
\end{align*}
which is equivalent to the fundamental identity \eqref{eq:FID}.

In summary, the result is a metric ternary algebra obeying the
fundamental identity and \eqref{eq:symmetry}, but the bracket is not
in general totally skewsymmetric.  This is easy to see because we saw
that for a metric 3-Lie algebra, the inner product on $\fg$ has split
signature, so from any $\fg$ which is, say, odd-dimensional one cannot
reconstruct a metric 3-Lie algebra.

By analogy with the relation between Lie and Leibniz algebras, let us
define a \textbf{(left) 3-Leibniz algebra} to be a vector space $V$
with a trilinear bracket $(x,y,z) \mapsto [x,y,z]$ satisfying the
fundamental identity \eqref{eq:FI}, whereas a metric 3-Leibniz algebra
possesses in addition a symmetric inner product obeying the metricity
axiom \eqref{eq:metricity}.  The metric ternary algebras just
constructed are special types of metric 3-Leibniz algebras where, in
addition, the symmetry condition \eqref{eq:symmetry} is satisfied.
These are precisely the metric 3-Leibniz algebras introduced by
Cherkis and Sämann in \cite{CherSaem}.

These 3-Leibniz algebras are such that the 3-bracket defines a linear
map $\Lambda^2 V \otimes V \to V$.  Decomposing
\begin{equation}
  \Lambda^2 V \otimes V = \Lambda^3 V \oplus V^{\yng(2,1)} \oplus V~,
\end{equation}
we see that there are two interesting limiting cases of these
algebras:
\begin{enumerate}
\item the metric 3-Lie algebras, for which the 3-bracket is totally
  skew-symmetric; and
\item the \textbf{metric Lie triple systems}, for which the 3-bracket
  obeys
  \begin{equation}
    \label{eq:LTS}
    [x,y,z] + [y,z,x] + [z,x,y] = 0~.
  \end{equation}
\end{enumerate}
A Lie triple system is such that $\fg \oplus V$ admits the structure
of a 2-graded Lie algebra (\emph{not} a Lie superalgebra) in such a
way that $[x,y,z] = [[x,y],z]$, whence \eqref{eq:LTS} becomes one
component of the Jacobi identity for $\fg \oplus V$.  The 2-graded Lie
algebra $\fg\oplus V$ is known as the \textbf{embedding Lie algebra}
of the Lie triple system.  The symmetry condition \eqref{eq:symmetry}
together with \eqref{eq:LTS} imply that the 4-tensor
$\left<[x,y,z],w\right>$ has the symmetries of an algebraic curvature
tensor.  This is not an accident: a metric Lie triple system is a
linear approximation to a (pseudo)riemannian symmetric space, in the
same way that a (metric) Lie algebra is a linear approximation to a
Lie group (possessing a bi-invariant metric), and the 4-tensor
$\left<[x,y,z],w\right>$ coincides with the Riemann curvature tensor
of the symmetric space.  It is a natural question (whose answer is not
known) whether there is any geometric object of which a 3-Lie algebra
is a linear approximation.

To summarise this lecture thus far, we started with a metric 3-Lie
algebra $V$ and we deconstructed it into a metric Lie algebra $\fg$
acting faithfully and orthogonally on $V$.  Conversely, starting from a
metric Lie algebra $\fg$ and a faithful orthogonal representation $V$,
we arrived at a strictly larger class of metric 3-Leibniz algebras,
including as special cases the metric 3-Lie algebras and the metric Lie
triple systems.  In general, of course, the general 3-Leibniz algebra in
this class is neither a Lie triple system nor 3-Lie.  It is an
interesting open problem to characterise the metric 3-Lie algebras
\emph{a priori} from their Lie-algebraic data.

One important remark is that the inner product on $\fg$ is an
important part of the data.  For example, let us take $\fg = \fso(4)$
and $V=\RR^4$ the fundamental representation.  If we take for the
inner product on $\fg$ (minus) the Killing form, then the resulting
3-Leibniz algebra is a Lie triple system with embedding Lie algebra
$\fso(5)$, whence this is the Lie triple system approximating linearly
the round 4-sphere thought of as the riemannian symmetric space
$\SO(5)/\SO(4)$.   Relative to an orthonormal basis $e_i$ for $\RR^4$,
the 3-brackets of this Lie triple system are given by
\begin{equation}
  [e_i,e_j,e_k] = \delta_{jk} e_i - \delta_{ik} e_j~.
\end{equation}
There is a one-parameter family of such 3-Leibniz algebras
interpolating between this Lie triple system and the simple 3-Lie
algebra $A_4$.  This is, in fact, the unique deformation of $A_4$
within the class of all 3-Leibniz algebras.  The deformation parameter
can be understood as parametrising the conformal classes of invariant
inner products on $\fso(4)$, which is the only part of the Lie
algebraic data which is not rigid.

\subsection{The Faulkner construction}
\label{sec:faulkn-constr}

We end this lecture by describing a general algebraic construction which
underlies the above deconstruction/reconstruction.

Let $\fg$ be a metric Lie algebra and let $V$ be a faithful
representation.  In contrast with the above discussion, we are not
assuming an inner product on $V$ for now.  Let $V^*$ denote the dual
representation, where if $X \in \fg$, $\alpha \in V^*$ and $v \in V$,
then
\begin{equation}
  (X \cdot \alpha)(v) = - \alpha(X \cdot v)~.
\end{equation}
Given $v \in V$ and $\alpha \in V^*$, we may define $\eD(v,\alpha) \in
\fg$ by
\begin{equation}
  \left(\eD(v,\alpha), X\right) = \alpha(X \cdot v) \qquad\text{for
    all $X\in\fg$.}
\end{equation}
It follows easily that if $X\in\fg$ is perpendicular to the image of
$\eD$, then it obeys $\alpha(X \cdot v) = 0$ for all $\alpha \in V^*$
and $v \in V$, which is equivalent to $X \cdot v = 0$ for all $v\in
V$.  Since $V$ is a faithful representation, this means that $X=0$.
This says that $\eD$ is a surjective linear map $V \otimes V^* \to
\fg$.  This allows us to define ``3-brackets'' mixing $V$ and $V^*$ by
\begin{equation}
  \begin{aligned}[m]
    V \times V^* \times V &\to V\\
    (v,\alpha,w) &\mapsto \eD(v,\alpha)\cdot w
  \end{aligned}
  \qquad\text{and}\qquad
  \begin{aligned}[m]
    V \times V^* \times V^* &\to V^*\\
    (v,\alpha,\beta) &\mapsto \eD(v,\alpha)\cdot \beta~,
  \end{aligned}
\end{equation}
which satisfy a version of the fundamental identity:
\begin{equation}
  [\eD(v,\alpha),\eD(w,\beta)] = \eD(\eD(v,\alpha)\cdot w,\beta) +
  \eD(w,\eD(v,\alpha)\cdot \beta)~.
\end{equation}

One could not call this a ternary algebra, however, because we do not
have a trilinear map on a single vector space: the ``brackets'' --- if
they could be called that --- involve both $V$ and $V^*$.  One way to
obtain an honest ternary algebra on $V$ is to identify $V$ and $V^*$
$\fg$-equivariantly, which requires the existence of a $\fg$-invariant
nondegenerate tensor in $V^* \otimes V^*$, e.g., an inner product.
There are seven elementary types of inner products on a vector space
$V$, but only three of them have a notion of signature.  We concentrate
on these because the inner product on $V$ appears in the kinetic terms
of the matter fields in the three-dimensional superconformal
Chern--Simons theories and manifest unitarity would dictate that we use
a positive-definite inner product.  The three types of positive-definite
inner product are real symmetric, complex hermitian and quaternionic
hermitian.  This means that $V$ is a real orthogonal, complex unitary or
quaternionic unitary representation of $\fg$, respectively.  In the real
and quaternionic cases, the inner product identifies $V$ and $V^*$ as
representations of $\fg$, whereas in the complex case it is $V^*$ and
$\overline V$ which are identified.  In this latter case, by restricting
scalars to the reals and in this way viewing $V$ as a real
representation (of twice the dimension) then we do have that again $V$
and $V^*$ are identified, whence the above general construction gives in
each case a metric 3-Leibniz algebra subject perhaps to further axioms.
The real case was already discussed above and as we saw gives rise to
the 3-Leibniz algebras of Cherkis and Sämann.  We will not discuss the
quaternionic case here, and simply refer the reader to \cite{Lie3Algs},
but in the next lecture, we will concentrate instead in the complex
case.

\section{The metric 3-Leibniz algebras of the $N{=}6$ theories}
\label{sec:metric-3-leibniz}

In this third and last lecture we discuss in detail the case of the
algebras underlying the $N{=}6$ theories.  This lecture too is based on
\cite{Lie3Algs}.

Aharony, Bergman, Jafferis and Maldacena (ABJM) \cite{MaldacenaBL}
constructed an $N{=}6$ superconformal Chern--Simons theory dual to
multiple M2-branes at an orbifold singularity.  Although written down in
a purely gauge-theoretic language, the model was reformulated in a
3-algebraic formalism by Bagger and Lambert in \cite{BL4}.  The
3-algebra underlying the simplest ABJM model is defined on the complex
vector space $V$ of $n\times n$ matrices with complex entries.  The
3-bracket is given by
\begin{equation}
  \label{eq:BL4ex}
  [x,y;z] = y z^\dagger x - x z^\dagger y~,
\end{equation}
for all $x,y,z \in V$ and where $z^\dagger$ is the hermitian adjoint of
$z$.  Although the 3-bracket is real trilinear, it is not complex
trilinear due to the presence of the hermitian adjoint.  Indeed, it is
evident from \eqref{eq:BL4ex} that the 3-bracket is complex linear in
the first two entries and complex antilinear in the third --- a fact
that is reflected in the notation for the 3-bracket.  It is similarly
evident that it is skewsymmetric in the first two entries:
\begin{equation}
  \label{eq:BL4skew}
  [x,y;z] = - [y,x;z]~.
\end{equation}
What may not be so evident is that, in addition, the 3-bracket satisfies
a version of the fundamental identity
\begin{equation}
  \label{eq:BL4FI}
  [[z,v;w],x;y] - [[z,x;y],v;w] - [z,[v,x;y];w] + [z,v;[w,y;x]] = 0~.
\end{equation}
Indeed, expanding each term, we get
\begin{align*}
    [[z,v;w],x;y] &= \phantom{+} x y^\dagger v w^\dagger z - x y^\dagger z w^\dagger v - v w^\dagger z y^\dagger x + z w^\dagger v y^\dagger x \\
  - [[z,x;y],v;w] &= - v w^\dagger x y^\dagger z + v w^\dagger z y^\dagger x + x y^\dagger z w^\dagger v - z y^\dagger x w^\dagger v \\
  - [z,[v,x;y];w] &= - x y^\dagger v w^\dagger z + v y^\dagger x w^\dagger z + z w^\dagger x y^\dagger v - z w^\dagger v y^\dagger x \\
    [z,v;[w,y;x]] &= \phantom{+} v w^\dagger x y^\dagger z - v y^\dagger x w^\dagger z - z w^\dagger x y^\dagger v + z y^\dagger x w^\dagger v~,
\end{align*}
and adding them we see that the 16 monomials do indeed cancel pairwise.  (The original fundamental identity in \cite{BL4} is different,
but as shown in \cite[Lemma~14]{Lie3Algs} they are equivalent for the class of algebras which obey \eqref{eq:BL4skew}.)
Finally, the vector space $V$ has a natural hermitian inner product:
\begin{equation}
\label{eq:BL4ip}
  h(x,y) = \Tr x y^\dagger~,
\end{equation}
satisfying the following compatibility condition with the 3-bracket:
\begin{equation}
  \label{eq:BL4symmetry}
  h([y,x; z], w) = h(y, [w,z; x])~.
\end{equation}
Indeed, expanding the left-hand side, we find
\begin{align*}
  h([y,x; z], w)  &= \Tr [y,x;z] w^\dagger\\
  &= \Tr (x z^\dagger y - y z^\dagger x) w^\dagger\\
  &= \Tr x z^\dagger y w^\dagger - \Tr y z^\dagger x w^\dagger\\
  &= \Tr y (w^\dagger x z^\dagger - z^\dagger x w^\dagger)\\
  &= \Tr y (z x^\dagger w - w x^\dagger z)^\dagger\\
  &= \Tr y [w,z;x]^\dagger\\
  &= h(y,[w,z;x])~.
\end{align*}

\subsection{Deconstructing the $N{=}6$ algebras}
\label{sec:deconstr-n=6-algebr}

A complex hermitian vector space $(V,h)$ with a bracket $(x,y,z) \mapsto
[x,y;z]$, complex linear in the first two entries and antilinear in the
third, satisfying properties \eqref{eq:BL4skew}, \eqref{eq:BL4FI} and
\eqref{eq:BL4symmetry}, can be deconstructed, as we did for metric 3-Lie
algebras in the second lecture, into a metric Lie algebra $\fg$ acting
on $V$ faithfully and preserving $h$.  Let us first consider the map $V
\times V \to \End V$ sending $(x,y)$ to $D(x,y):= [-,x;y]$.  Notice that
this map is sesquilinear: complex linear in the first entry and complex
antilinear in the second.  In terms of this map, the fundamental
identity \eqref{eq:BL4FI} can be written as
\begin{equation}
  \label{eq:BL4FID}
  [D(x,y),D(v,w)] =  D(D(x,y)\cdot v,w) - D(v,D(y,x)\cdot w)
\end{equation}
and the symmetry condition \eqref{eq:BL4symmetry} can be written as
\begin{equation}
  \label{eq:BL4symmetryD}
  h(D(x,z)\cdot y, w) = h(y, D(z,x) \cdot w)~.
\end{equation}
Equation \eqref{eq:BL4FID} says that the image of $D$ is a
\emph{complex} Lie subalgebra of $\fgl(V)$ denoted $\fg_{\CC}$, as it
will be seen to be the complexification of a real Lie algebra $\fg$.

To understand what $\fg$ might be, we use the symmetry condition, which
we would like to reinterpret as a unitarity condition.  Of course, a
complex Lie algebra cannot leave a hermitian inner product invariant:
instead one has the condition
\begin{equation}
  h(X \cdot u, v) = - h(u, \overline X \cdot v)~,
\end{equation}
for all $X \in \fg_{\CC}$ and where $X \mapsto \overline X$ is a
conjugation on the Lie algebra, whose fixed point set is the real Lie
algebra $\fg$ we are after.  From equation \eqref{eq:BL4symmetryD}, we
see that
\begin{equation}
  \overline{D(x,z)} = - D(z,x)~,
\end{equation}
whence $\fg$ is spanned by the real parts
\begin{equation}
  \label{eq:BL4E}
  E(x,y):= D(x,y) + \overline{D(x,y)} = D(x,y) - D(y,x)~.
\end{equation}
An easy consequence of the fundamental identity \eqref{eq:BL4FID} is
that
\begin{equation}
  \label{eq:BL4FIE}
  [E(x,y),E(v,w)] =  E(E(x,y)\cdot v,w) + E(v,E(x,y)\cdot w)~.
\end{equation}
Indeed, expanding the left-hand side and using \eqref{eq:BL4E},
\eqref{eq:BL4FID} and  \eqref{eq:BL4E} again, we obtain
\begin{align*}
  [E(x,y),E(v,w)] &=  [D(x,y)-D(y,x),D(v,w)-D(w,v)]\\
  &= D(D(x,y)\cdot v, w) - D(v, D(y,x) \cdot w) - D(D(y,x)\cdot v, w)\\
  &\quad + D(v, D(x,y) \cdot w) - D(D(x,y)\cdot w, v) + D(w, D(y,x)
  \cdot v) \\
  &\quad + D(D(y,x)\cdot w, v) - D(w, D(x,y) \cdot v) \\
  &= E(E(x,y) \cdot v, w) + E(v,E(x,y)\cdot w)~.
\end{align*}

As in the second lecture, it is easy to show that $\fg$ is metric,
relative to the inner product
\begin{equation}
  \label{eq:BL4GIP}
  \left(E(x,y), E(u,v)\right) = \Re\, h(E(x,y)\cdot u, v)~,
\end{equation}
which can again be shown to be symmetric, nondegenerate and
$\fg$-invariant.  Let us prove each property in turn.  To prove symmetry
we simply calculate:
  \begin{align*}
    h(E(x,y)\cdot u, v) &= h(D(x,y)\cdot u, v)- h(D(y,x)\cdot u, v)\\
    &= h([u,x;y], v) - h([u,y;x], v)\\
    &= -h([x,u;y],v) + h([y,u;x], v) && \text{using \eqref{eq:BL4skew}}\\
    &= -h(x,[v,y;u]) + h(y,[v,x;u])  &&\text{using \eqref{eq:BL4symmetry}}\\
    &= h(x,[y,v;u]) - h(y,[x,v;u])  &&\text{using \eqref{eq:BL4skew} again}\\
    &= h(x,D(v,u)\cdot y) - h(y, D(v,u)\cdot x)\\
    &= h(D(u,v)\cdot x, y) - \overline{h(D(v,u)\cdot x, y)}~,
  \end{align*}
  whence taking real parts we find
  \begin{equation*}
    \Re h(E(x,y)\cdot u, v) = \Re h (E(u,v)\cdot x, y)~.
  \end{equation*}
  To prove nondegeneracy, let us assume that some linear combination
  $X:= \sum_i E(x_i, y_i)$ is orthogonal to all $E(u,v)$, so that
  \begin{equation*}
    \Re h(X \cdot u, v) = 0 \qquad\text{for all $u,v\in V$.}
  \end{equation*}
  Now, since $h$ is nondegenerate, so is $\Re h$ because $\Im h(x,y) =
  \Re h(-i x, y)$, whence this means that $X\cdot u = 0$ for all $u$,
  showing that the endomorphism $X =0$.  Finally, we show that it is
  $\fg$-invariant.  Using \eqref{eq:BL4FIE}, we find
  \begin{align*}
    \left(E(z,w),[E(x,y),E(u,v)]\right) &= \left(E(z,w), E(E(x,y)\cdot
      u, v) + E(u, E(x,y)\cdot v)\right)\\
    &= \Re h(E(z,w)\cdot E(x,y)\cdot u, v) + \Re h( E(z,w)\cdot u,
    E(x,y) \cdot v)\\
    &= \Re h(E(z,w)\cdot E(x,y)\cdot u, v) - \Re h( E(x,y)\cdot
    E(z,w)\cdot u, v)\\
    &= \Re h([E(z,w), E(x,y)]\cdot u, v) \\
    &= \left([E(z,w), E(x,y)], E(u,v)\right)~,
  \end{align*}
which is the ad-invariance of the inner product on $\fg$.

For example, for the algebra of equation \eqref{eq:BL4ex}, one finds
\begin{equation}
  D(y,z) \cdot x =  y z^\dagger x - x z^\dagger y \implies E(y,z) \cdot
  x = (y z^\dagger - z y^\dagger) x + x (y^\dagger z - z^\dagger y)~.
\end{equation}
The $n\times n$ matrices $y z^\dagger - z y^\dagger$ and $y^\dagger z -
z^\dagger y$ are skewhermitian, whence in $\fu(n)$.  Their traces sum to
zero, whence only the $\fsu(n)$ components act effectively on $V$.  In
other words, $\fg = \fsu(n) \oplus \fsu(n)$ and $V$ is the bifundamental
representation $(\boldsymbol{n},\overline{\boldsymbol{n}})$.  The inner
product \eqref{eq:BL4GIP} on $\fg$ has split signature, being given by
the difference of the traces in the fundamental representations:
\begin{equation}
  \left(X_L\oplus X_R, Y_L \oplus Y_R\right)= \Tr X_L Y_L - \Tr X_R Y_R~,
\end{equation}
for all $X_L,X_R,Y_L,Y_R \in \fsu(n)$.

\subsection{Reconstructing the $N{=}6$ algebras}
\label{sec:reconstr-n=6-algebr}

Conversely, let us start with a metric Lie algebra $\fg$ with inner
product $\left(-,-\right)$ and a faithful complex unitary representation
$(V,h)$.  (In our somewhat unusual conventions, the hermitian inner
product is complex antilinear in the second argument.) We will
reconstruct a 3-bracket $(x,y,z) \mapsto [x,y;z]$ on $V$, complex linear
in the first two entries and antilinear in the third, obeying both the
fundamental identity \eqref{eq:BL4FI} and the symmetry condition
\eqref{eq:BL4symmetry}, but not in general the skewsymmetry condition
\eqref{eq:BL4skew}, which lands us in a similar situation as with the
metric 3-Lie algebras.  This then prompts us to ask how to characterise
those 3-algebras which obey \eqref{eq:BL4skew} and we will see that they
are characterised in terms of certain kinds of metric Lie superalgebras,
in agreement with an observation in \cite{SchnablTachikawa} based on
\cite{3Lee,GaiottoWitten}.

Let $\fg_{\CC}$ denote the complexification of $\fg$.  We extend the
inner product complex bilinearly in such a way that $\fg_{\CC}$ becomes
a complex metric Lie algebra.  Similarly we extend the action of $\fg$
on $V$ to an action of $\fg_{\CC}$, using the fact that $V$ is already a
complex vector space.  This action remains faithful, but it is no longer
unitary.  Instead, we have
\begin{equation}
  \label{eq:pre-unitarity-C}
  h(\XX \cdot v, w) = - h(v, \XXbar\cdot w)~,
\end{equation}
for all $v,w\in V$ and $\XX \in \fg_{\CC}$ with $\XXbar$ its complex
conjugate.  Given $v,w \in V$, we define $D(v,w) \in \fg_{\CC}$ by
transposing the action of $\fg_{\CC}$ on $V$.  Explicitly, we have
\begin{equation}
  \label{eq:D-C}
  \left(D(v,w), \XX\right) = h(\XX\cdot v, w)~,
\end{equation}
which shows that $D(v,w)$ is complex linear in $v$, but complex
antilinear in $w$, whence it defines a sesquilinear map $D: V \times V
\to \fg_{\CC}$.  As before, we see that image of $D$ is all of
$\fg_{\CC}$, since if $\XX$ is perpendicular to the image of $D$, it
must annihilate all $v\in V$, and since the representation is faithful,
then $\XX=0$.

Complex conjugating \eqref{eq:D-C}, we find
\begin{align*}
  (\overline{D(v,w)},\XXbar) &= \overline{h(\XX\cdot
    v, w)} && \text{by \eqref{eq:D-C}}\\
  &= h(w,\XX\cdot v) &&\text{since $h$ is hermitian}\\
  &= - h(\XXbar \cdot w, v) &&\text{by \eqref{eq:pre-unitarity-C}}\\
  &= - (D(w,v),\XXbar)~,
\end{align*}
whence
\begin{equation}
  \label{eq:Dbar}
  \overline{D(v,w)} = - D(w,v)~.
\end{equation}

Now let $\XX \in \fg_{\CC}$ and $v,w\in V$.  Then for all $\YY \in
\fg_{\CC}$ we have,
\begin{align*}
  \left([D(v,w),\XX],\YY\right) &= \left(D(v,w),[\XX,\YY]\right)
  &&\text{since $(-,-)$ is invariant}\\
  &= h([\XX,\YY]\cdot v, w) &&\text{by \eqref{eq:D-C}}\\
  &= h(\XX \cdot \YY \cdot v, w) - h(\YY \cdot \XX \cdot v, w) &&
  \text{since $V$ is a representation}\\
  &= - h(\YY \cdot v, \XXbar \cdot w) - h(\YY \cdot \XX \cdot v, w) && \text{by \eqref{eq:pre-unitarity-C}}\\
  &= - \left(D(v,\XXbar \cdot w), \YY \right) - \left(D(\XX\cdot v, w),\YY\right) && \text{again by \eqref{eq:D-C}}~,
\end{align*}
whence abstracting $\YY$,
\begin{equation}
  \label{eq:ideal-C}
  [\XX,D(v,w)] = D(\XX\cdot v, w) + D(v, \XXbar\cdot w)~.
\end{equation}
Substituting $\XX = D(x,y)$, whence $\XXbar = - D(y,x)$, we obtain
equation \eqref{eq:BL4FID}.  This in turn is equivalent to the
fundamental identity \eqref{eq:BL4FI} for the ``2$\half$-bracket'' $V \times V
\times V \to V$ defined by
\begin{equation*}
  [x,y;z] := D(y,z) \cdot x~.
\end{equation*}
In terms of this bracket, equation \eqref{eq:pre-unitarity-C} becomes
\eqref{eq:BL4symmetry}, whereas the symmetry of the inner product on
$\fg_{\CC}$, applied to $\left(D(x,y),D(v,w)\right)$ becomes
\begin{equation}
  h([x,v;w],y) = h([v,x;y],w)~.
\end{equation}

However the condition \eqref{eq:BL4skew} does not follow from the
construction and must be imposed by hand.  This prompts the question of
how to characterise the data $\fg$, $\left(-,-\right)$ and $(V,h)$ such
that condition \eqref{eq:BL4skew} is satisfied.  We still don't know how
to do this a priori, but we can nevertheless characterise those algebras
which do in terms of metric Lie superalgebras.  The details appear in
\cite[Section~3.3]{Lie3Algs}.  To summarise, condition
\eqref{eq:BL4skew} turns out to be one component of the Jacobi identity
in a complex Lie superalgebra with underlying vector space $\fg_{\CC}
\oplus (V \oplus \overline V)$ and whose only nonvanishing odd-odd
bracket is given by $D: V \otimes V \to \fg_{\CC}$, which being
sesquilinear means that $[V,V]=[\overline V,\overline V] = 0$.
Furthermore this complex Lie superalgebra is the complexification of a
metric real Lie superalgebra with underlying vector space $\fg \oplus
\rf{V}$, where $\rf{V}$ is the real vector space obtained from $V$ by
restricting scalars to $\RR$ or, equivalently, $\rf{V}\otimes \CC = V
\oplus \overline{V}$.

In summary, there is a one-to-one correspondence between the metric
3-algebras in the Bagger--Lambert description of the $N{=}6$ theories of
ABJM-type and metric real Lie superalgebras $\fg \oplus \rf{V}$ with $V$
a complex unitary representation of $\fg$ whose only nonvanishing
odd-odd brackets are of mixed type.  For the example \eqref{eq:BL4ex},
the corresponding Lie superalgebra is the real form $\fpsu(n|n)$ of the
simple Lie superalgebra $A(n-1,n-1)$.

The emerging picture is thus the following: three-dimensional
Chern--Simons + matter theories admit a formulation in terms of metric
ternary algebras, which can be constructed from a metric Lie algebra and
a faithful unitary representation.  The generic ternary algebras
obtained in this way should correspond to theories with $N\leq 3$
supersymmetry, whereas for $N\geq 4$ supersymmetry, we need to
specialise to ternary algebras obeying additional symmetry conditions,
as we have seen for the $N{=}6$ and the $N{=}8$ theories in this and the
previous lectures.  The precise dictionary between the amount of
supersymmetry and the type of ternary algebra is the subject of a
forthcoming publication.

\section*{Acknowledgments}

These notes are based on lectures delivered at the Universitat de
València in October 2008 and in Nordita in November 2008 as part of
the workshop \emph{Geometrical aspects of String Theory} organised by
Ulf Lindström and Maxim Zabzine, to whom I am grateful for the
invitation to participate.  The results described in these notes are
the result of a very enjoyable ongoing collaboration with Paul de
Medeiros, Elena Méndez-Escobar and Patricia Ritter, whom I take
pleasure in thanking.

Finally, these notes were prepared while on sabbatical at the
Universitat de València, supported under research grant FIS2008-01980.
I am grateful to José de Azcárraga for making this visit possible.

\bibliographystyle{utphys}
\bibliography{AdS,AdS3,ESYM,Sugra,Geometry,Algebra}

\end{document}